\documentstyle[twoside,fleqn,espcrc2,epsf,epsfig]{article}

\newcommand{\AmS}{{\protect\the\textfont2
A\kern-.1667em\lower.5ex\hbox{M}\kern-.125emS}}

\title{Strong randomness of off-diagonal gluon phases and off-diagonal 
gluon mass in the maximally abelian gauge in 
QCD\thanks{Talk presented by H.~Suganuma}}
\author{H.~Suganuma\address[TIT]{
\vspace{-0.05cm}
Faculty of Science, 
Tokyo Institute of Technology, 
Ohokayama 2-12-1, Meguro, Tokyo 152-8551, Japan}
K.~Amemiya\address[AAS]{
\vspace{-0.05cm}
Advanced Algorithm and Systems, 
Ebisu 1-13-6, Shibuya, Tokyo 150-0013, Japan}, 
H.~Ichie\address[HB]{
\vspace{-0.05cm}
Humboldt Univ. zu Berlin, 
Institut f\"ur Physik, Invalidenstrasse 110, D-10115 Berlin, Germany},
N.~Ishii\address[RIKEN]{
\vspace{-0.05cm}
The Institute of Physical and Chemical Research (RIKEN), 
Hirosawa 2-1, Wako 351-0198, Japan},
H.~Matsufuru\address[YIPT]{
\vspace{-0.05cm}
Yukawa Institute for Theoretical Physics (YITP), 
Kyoto University, Sakyo, Kyoto 606-8502, Japan}
and 
T.T.~Takahashi\address{
\vspace{-0.05cm}
Research Center for Nuclear Physics, 
Osaka University, Mihogaoka 10-1, Ibaraki 567-0047, Japan}
}
\begin{document}
\begin{abstract} 
We study abelianization of QCD in the maximally abelian (MA) gauge. 
In the MA gauge, the off-diagonal gluon amplitude is 
strongly suppressed, and then the off-diagonal gluon phase shows 
strong randomness, which leads to a large off-diagonal gluon mass.
Using lattice QCD, we find a large effective off-diagonal gluon mass 
in the MA gauge: $M_{\rm off} \simeq 1.2 {\rm GeV}$ in SU(2) QCD,  
$M_{\rm off} \simeq 1.1 {\rm GeV}$ in SU(3) QCD. 
Due to the large off-diagonal gluon mass in the MA gauge, 
infrared QCD is well abelianized like nonabelian Higgs theories. 
We investigate the inter-monopole potential and the dual gluon field 
$B_\mu$ in the MA gauge, and find longitudinal magnetic screening 
with $m_B \simeq$ 0.5 GeV in the infrared region, 
which indicates the dual Higgs mechanism by monopole condensation.
We propose a gauge invariant description of the MA projection 
by introducing the ``gluonic Higgs scalar field''.
\vspace{-0.275cm}
\end{abstract} 

\maketitle

\section{Strong Randomness of Off-diagonal Gluon Phase  
and Large Mass of Off-diagonal Gluons in MA Gauge}

In Euclidean QCD, the maximally abelian (MA) gauge is 
defined so as to minimize the total amount of the 
off-diagonal gluons \cite{CONF2000,IS9900,SAIT00,SITA98}
\begin{equation}
\vspace{-0.3cm}
R_{\rm off} [A_\mu ( \cdot )] \equiv \int d^4x \ {\rm tr}
\left\{ 
[\hat D_\mu ,\vec H][\hat D_\mu ,\vec H]^\dagger 
\right\} 
\vspace{-0.3cm}
\end{equation}
by the SU($N_c$) gauge transformation, from which 
the local MA gauge condition is easily derived as 
$[\vec H, [\hat D_\mu , [\hat D_\mu , \vec H]]]=0$.

In SU(2) lattice QCD, 
we find two remarkable features of the off-diagonal gluon  
$A_\mu^\pm \equiv \frac1{\sqrt{2}}(A_\mu^1 \pm i A_\mu^2)
=e^{\pm i \chi_\mu(x)}|A_\mu^\pm(x)|$ in the MA gauge 
\cite{CONF2000,IS9900,SAIT00}. 
\begin{enumerate}
\item
The off-diagonal gluon amplitude $|A_\mu^{\pm}(x)|$ 
is strongly suppressed by SU($N_c$) gauge transformation in the MA gauge.
\vspace{-0.1cm}
\item
The off-diagonal gluon phase $\chi_\mu(x)$ tends 
to be random, because $\chi_\mu(x)$ is not 
constrained by MA gauge fixing at all, 
and only the constraint from the QCD action is weak 
due to a small accompanying factor $|A_\mu^\pm|$.
\vspace{-0.1cm}
\end{enumerate}
We investigate 
$\Delta \chi \equiv |\chi_\mu(s)-\chi_\mu(s+\hat \nu)| ({\rm mod } \pi)$ 
in the MA gauge with U(1)$_3$ Landau gauge fixing. 
If the off-diagonal gluon phase  $\chi_\mu(x)$ is a continuum variable,  
as the lattice spacing $a$ goes to 0, 
$\Delta \chi \simeq a |\partial_\nu \chi_\mu|$  
goes to zero, and the probability distribution 
$P(\Delta \chi)$ approaches to $\delta(\Delta \chi)$. 
However, $P(\Delta \chi)$ is almost flat 
independently of $a$ or $\beta$, which indicates  
{\it strong randomness of the off-diagonal gluon phase} 
$\chi_\mu(x)$ in the MA gauge. 

Remarkably,  
{\it strong randomness of off-diagonal gluon phases} 
in the MA gauge leads to   
{\it rapid reduction of off-diagonal gluon correlations}   
\cite{CONF2000} as 
\begin{eqnarray}
\langle A_\mu^+(x) A_\nu^-(y) 
\rangle \!\!\!\!\!&=&\!\!\!\!\! 
\langle |A_\mu^+(x)A_\nu^-(y)| e^{i\{\chi_\mu(x)-\chi_\nu(y)\}} 
\rangle \nonumber \\ 
\!\!\!\!\!&\sim&\!\!\!\!\! \langle |A_\mu^\pm(x)|^2 \rangle_{\rm MA} 
\delta_{\mu\nu}\delta^4(x-y), 
\vspace{-0.1cm}
\end{eqnarray}
which means the infinitely large mass of off-diagonal gluons.
Since the real off-diagonal gluon phases are not complete but 
approximate random phases even in the MA gauge, 
the off-diagonal gluon mass $M_{\rm off}$ would be large but finite. 

\section{Large Mass Generation of Off-diagonal Gluons in MA Gauge : 
Essence of Infrared Abelianization of QCD}

We study the Euclidean gluon propagator 
$G_{\mu \nu }^{ab} (x-y) \equiv \langle A_\mu ^a(x)A_\nu ^b(y)\rangle$ 
($a,b =1,2,.., N_c^2-1$) and the off-diagonal gluon mass $M_{\rm off}$ 
in the MA gauge 
\cite{CONF2000,SAIT00,SITA98,AS99}
using SU($N_c$) lattice QCD with $N_c=2,3$.
As for the residual abelian gauge symmetry, 
we take ${\rm U(1)}_3 (\times {\rm U(1)}_8) $ Landau gauge, 
to extract most continuous gluon configuration under 
the MA gauge constraint, for the comparison with the continuum theory.
The continuum gluon field $A_\mu^a(x)$ is derived from 
the link variable as 
$U_\mu(s)={\rm exp}\{iaeA_\mu^a(s) T^a \}$.

We show the scalar-type gluon propagators
$G_{\mu \mu}^{a}(r) 
\equiv \langle A_\mu^{a}(x)A_\mu^{a}(y)\rangle$ ($a=1,2,.., N_c^2-1$), 
which depend only on the four-dimensional Euclidean 
distance $r \equiv \sqrt{(x_\mu- y_\mu)^2}$.
The four-dimensional Euclidean propagator of the 
massive vector boson with the mass $M$ takes a 
Yukawa-type asymptotic form as 
\begin{equation}
G_{\mu\mu}(r) \simeq \frac3{4\pi^2} \frac{M}{r} K_1(Mr)
\simeq \frac{3M^{1/2}}{2(2\pi)^{3/2}}\frac{e^{-Mr}}{r^{3/2}}.
\end{equation}
From the slope analysis of the lattice QCD data of 
$\ln\{r^{3/2}G_{\mu\mu}(r)\}$ with $r \ge 0.2 {\rm fm}$,  
we obtain the off-diagonal gluon mass in the MA gauge as follows. 
\begin{enumerate}
\item
$M_{\rm off} \simeq 1.2~{\rm GeV}$ 
in SU(2) lattice QCD\footnote{
From the mass measurement with 
the zero-momentum projection 
in SU(2) lattice QCD ($2.3 \le \beta \le 2.35$,   
$16^3\times 32$, $12^3\times 24$) we find again 
$M_{\rm off} \simeq 1.2 {\rm GeV}$ 
in the MA gauge. 
}
with $2.2 \le \beta \le 2.4$ 
and $12^3 \times 24$, $16^4$, $20^4$ (Fig.1).
\item
$M_{\rm off} \simeq 1.1~{\rm GeV}$ 
in SU(3) lattice QCD with $\beta=5.7$ ($a \simeq $ 0.19fm) 
and $12^3 \times 24$ (Fig.2). 
\end{enumerate}
In SU(3) QCD, 
the two diagonal gluon propagators, 
$G_{\mu\mu}^{3}$ and $G_{\mu\mu}^{8}$, show the same 
large distance propagation, and 
the three off-diagonal gluon propagators 
$G_{\mu\mu}^{+-(i,j)} \equiv \langle 
A_\mu^{+(i,j)}(x)A_\mu^{-(i,j)}(y) \rangle$ 
with $A_\mu^{\pm(i,j)} \!\!\!\equiv\!\!\! \frac1{\sqrt{2}}
(A_\mu^i \pm i A_\mu^j)$ with $(i,j)=(1,2),$ 
$(4,5),(6,7)$ show the same massive behavior.

\begin{figure}[htb]
\vspace{-0.56cm}
\begin{center}
\epsfig{figure=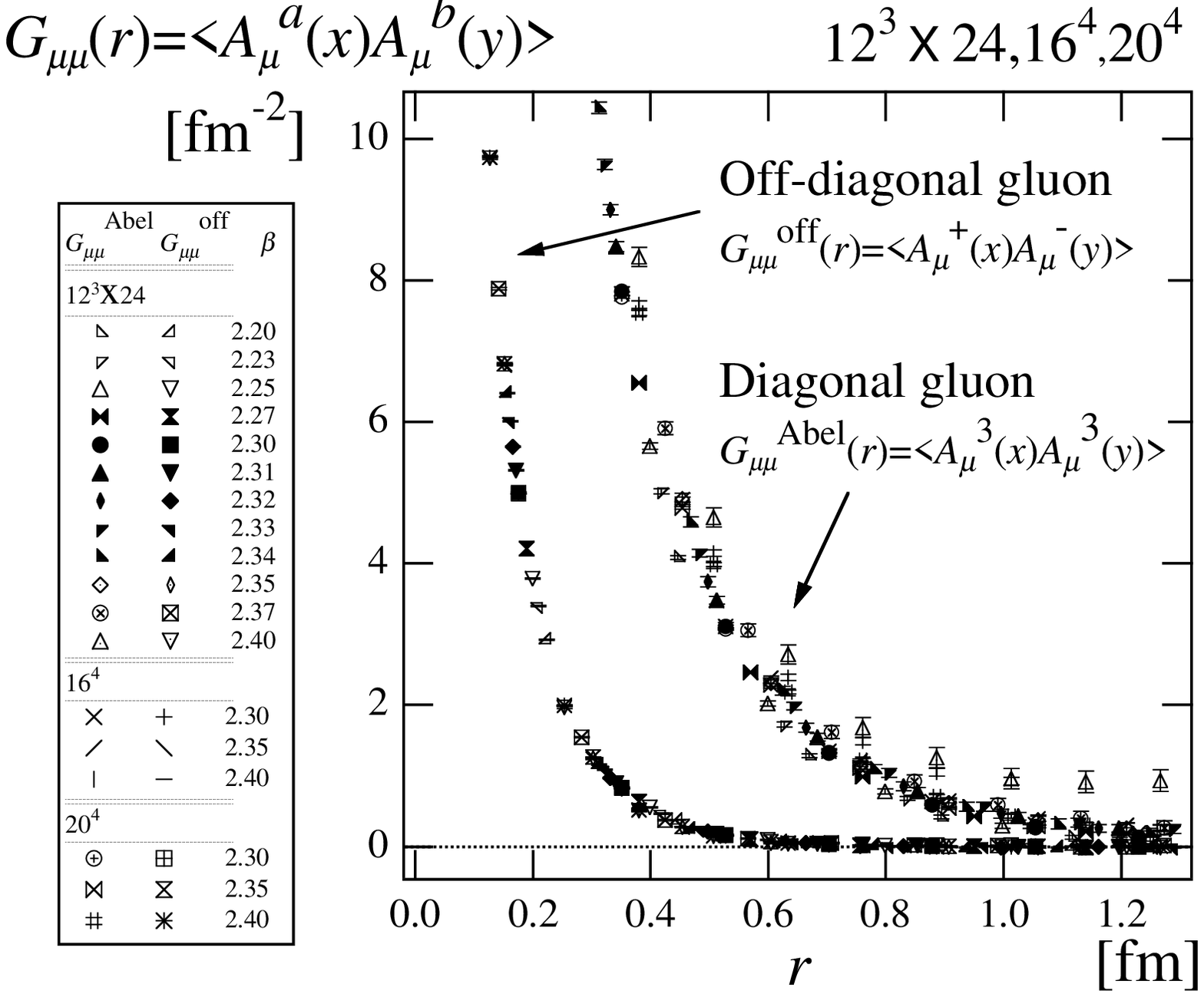,height=4.81cm} 
\epsfig{figure=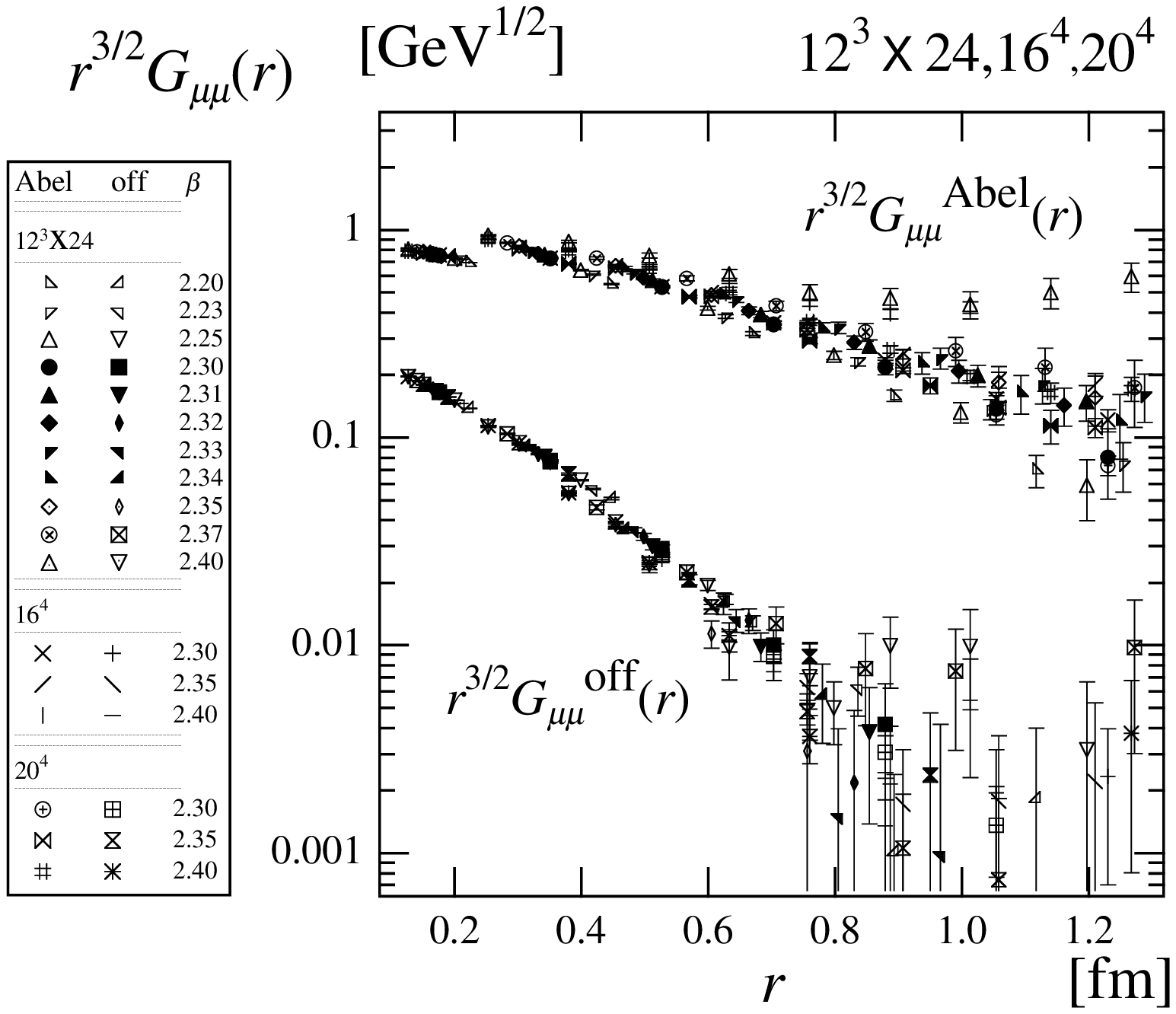,height=4.81cm}
\vspace{-1.5cm}
\caption{(a) The scalar-type gluon propagator 
$G_{\mu \mu }^a(r)$ v.s. 4-dim.~distance $r$ 
in the MA gauge in SU(2) lattice QCD with  
$2.2 \le \beta \le 2.4$, $12^3 \times 24$, $16^4$, $20^4$. 
(b) The logarithmic plot of $r^{3/2} G_{\mu \mu}^a(r)$. 
}
\end{center} 
\vspace{-1.27cm}
\end{figure}

\begin{figure}[htb]
\vspace{-0.35cm}
\begin{center}
\epsfig{figure=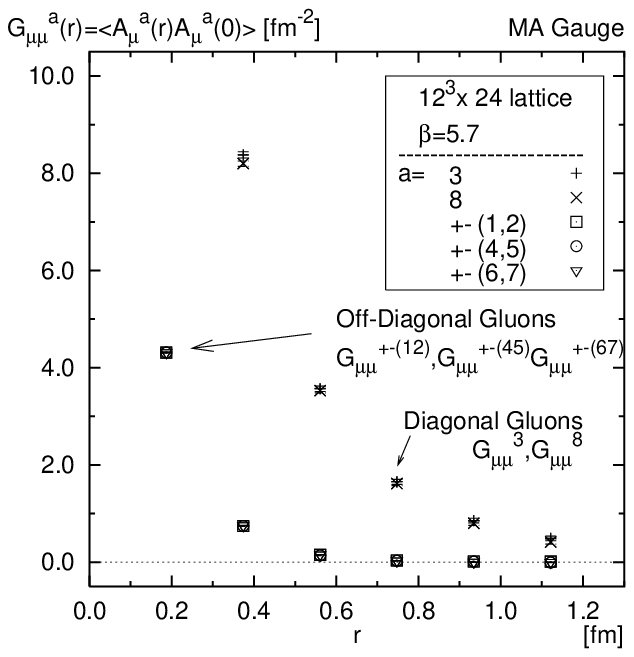,height=4.85cm}
\epsfig{figure=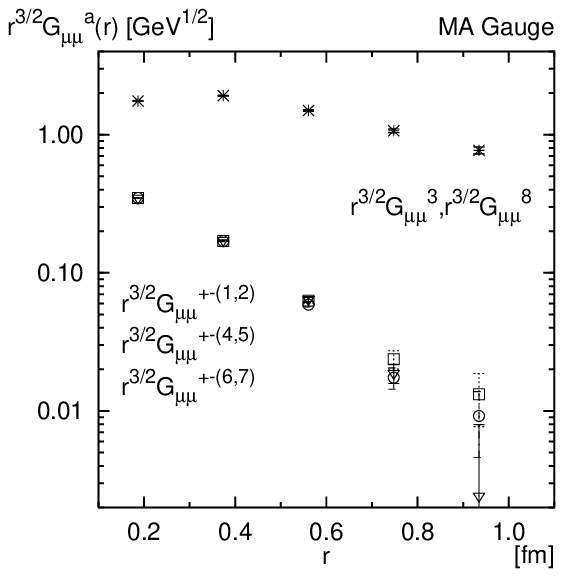,height=4.85cm}
\vspace{-1.5cm}
\caption{
(a) The scalar-type gluon propagator 
$G_{\mu \mu }^a(r)$ v.s. the four-dimensional 
distance $r$ in the MA gauge 
with ${\rm U(1)}_3 \times {\rm U(1)}_8$ Landau gauge fixing 
in SU(3) lattice QCD with
$\beta=5.7$ and $12^3 \times 24$.
(b) The logarithmic plot of $r^{3/2} G_{\mu \mu}^a(r)$. 
}
\end{center}
\vspace{-1.13cm}
\end{figure}

To conclude, both in SU(2) and SU(3) QCD, 
the {\it off-diagonal gluon acquires a 
large effective mass  
$M_{\rm off} \simeq 1 {\rm GeV}$ in the MA gauge}, which is 
{\it essence of infrared abelian dominance} 
\cite{CONF2000,SAIT00,SITA98,AS99,EI82,M95,K9899}.

\section{Inter Monopole Potential, Longitudinal Magnetic Screening,   
Infrared Monopole Condensation and Monopole Structure}
\vspace{-0.13cm}

Using SU(2) lattice QCD, we study 
the {\it inter-monopole potential} and 
the {\it dual gluon propagator} 
in the monopole part in the MA gauge,  
and show {\it longitudinal magnetic screening} in the infrared region, 
as a direct evidence of 
the {\it dual Higgs mechanism by monopole condensation} 
\cite{N74,tH81}.
The dual gluon mass is estimated as $m_B \simeq$ 0.5 GeV  
\cite{CONF2000,SAIT00,SITA98}.
Then, lattice QCD in the MA gauge exhibits 
{\it infrared abelian dominance} and 
{\it infrared monopole condensation}, 
which lead to 
the dual Ginzburg-Landau (DGL) theory \cite{SST95} for infrared QCD.

Using SU(2) lattice QCD in the MA gauge, 
we find also the {\it monopole structure 
relating to a large amount of off-diagonal gluons 
around its center} like the 't~Hooft-Polyakov monopole.  
At a large scale, off-diagonal gluons 
inside monopoles become invisible, 
and monopoles can be regarded as point-like Dirac monopoles 
\cite{CONF2000,IS9900,SAIT00,SITA98}. 

\section{
Gluonic Higgs and Gauge Invariant Description of MA Projection}
\vspace{-0.1cm}

We propose a {\it gauge invariant description 
of the MA projection in QCD} 
\cite{CONF2000,IS9900} 
by introducing a {\it ``gluonic Higgs scalar field''} 
$\vec \phi(x) \equiv \Omega(x) \vec H \Omega^\dagger(x)$ 
with $\Omega(x) \in {\rm SU}(N_c)$ 
so as to minimize 
\begin{equation}
\vspace{-0.55cm}
R[\vec \phi(\cdot)] \equiv \int d^4x \ {\rm tr} 
\left\{[\hat D_\mu, \vec \phi(x)][\hat D_\mu, \vec \phi(x)]^\dagger \right\}
\vspace{-0.55cm}
\end{equation}
for an arbitrary given gluon field $\{A_\mu(x)\}$. 

The gluonic Higgs scalar $\vec \phi(x)$ physically corresponds to 
a ``color-direction'' of the nonabelian gauge connection $\hat D_\mu$ 
averaged over $\mu$ at each $x$. 

Through the projection along $\vec \phi(x)$, 
we can extract the abelian U(1)$^{N_c-1}$ sub-gauge-manifold 
which is most close to the original SU($N_c$) gauge manifold. 
This projection is manifestly gauge invariant, and 
is mathematically equivalent to the ordinary 
MA projection \cite{CONF2000,IS9900}. 
In this description, we find 
{\it infrared relevance of the gluon mode along 
the color-direction $\vec \phi(x)$} \cite{CONF2000,IS9900,C8000}, 
corresponding to infrared abelian dominance in the MA gauge.

Similar to $\hat D_\mu$, the gluonic Higgs scalar $\vec \phi(x)$ 
obeys the adjoint gauge transformation, 
and $\vec \phi(x)$ is diagonalized in the MA gauge. 
Then, {\it monopoles appear at the hedgehog singularities of $\vec \phi(x)$,}  
when the gluon field is continuous as in the SU(2) Landau gauge 
as shown in Fig.3 \cite{CONF2000,IS9900}.

\begin{figure}[htb]
\vspace{-0.8cm}
\begin{center}
\epsfig{figure=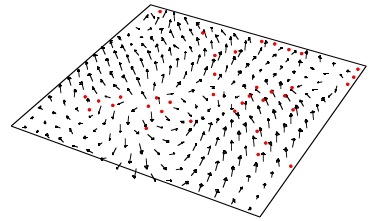,height=0.38\columnwidth}
\vspace{-0.8cm}
\caption{
The correlation between the gluonic Higgs scalar field 
$\phi(x)=\phi^a(x)\frac{\tau^a}{2}$ and monopoles (dots)  
in the SU(2) Landau gauge 
in SU(2) lattice QCD with $\beta=2.4$ and $16^4$.
The arrow denotes the SU(2) color direction of 
$(\phi^1(x),\phi^2(x),\phi^3(x))$.
The monopole appears at the hedgehog singularity of 
the gluonic Higgs scalar $\phi(x)$. 
}
\end{center}
\end{figure}

\end{document}